\documentclass[11pt,twoside]{article} 
\usepackage{adassconf}

\newcommand{\figref}[1]{Fig.~\ref{#1}}

\newcommand{\pu}[2]{\ensuremath{#1\,\mbox{#2}}}

\paperID{O10.1}
\begin{document}

\nopagebreak
\title{Gaia Data Processing Architecture}

\author{W.~O'Mullane, U.~Lammers} 
\affil{European Space Astronomy Centre, Madrid, Spain}
\author{C.~Bailer-Jones}
\affil{Max Planck Institut f\"ur Astronomie, Heidelberg, Germany}
\author{U.~Bastian}
\affil{Astronomisches Recheninstitut (ARI), Heidelberg, Germany}
\author{A.G.A.~Brown}
\affil{Sterrewacht Leiden, Leiden University, Leiden, The Netherlands}
\author{R.~Drimmel}
\affil{Osservatorio Astronomico di Torino, Torino, Italy}
\author{L.~Eyer}
\affil{Observatoire de Gen\`eve, Sauverny, Switzerland}
\author{C.~Huc}
\affil{Centre National d'Etudes Spatiales (CNES), Toulouse, France}
\author{D.~Katz}
\affil{Meudon Observatory, Paris, France}
\author{L.~Lindegren}
\affil{Lund Observatory, Lund, Sweden}
\author{D.~Pourbaix}
\affil{Universit\'e Libre de Bruxelles, Belgium}
\author{X.~Luri, J.~Torra}
\affil{University of Barcelona, Barcelona, Spain}
\author{F.~Mignard}
\affil{Observatoire de la C\^ote d'Azur, Nice, France}
\author{F.~van Leeuwen}
\affil{Institute of Astronomy, Cambridge, England}

\contact{William O'Mullane}
\email{William.OMullane@sciops.esa.int}

\paindex{ O'Mullane, W.}
\aindex{ Lammers, U.}
\aindex{ Bailer-Jones, C.}
\aindex{Bastian, U.}
\aindex{Brown, A.G.A.}
\aindex{Drimmel, R.}
\aindex{Eyer, L.}
\aindex{Huc, C.}
\aindex{Katz,D.}
\aindex{Lindegren,L.}
\aindex{Pourbaix, D.}
\aindex{Luri, X.}
\aindex{Mignard,F.}
\aindex{Torra, J.}
\aindex{van Leeuwen, F.}


\begin{abstract}
Gaia is ESA's ambitious space astrometry mission with a main objective to
astrometrically and spectro-photometrically map not less than 1000 million
celestial objects in our galaxy with unprecedented accuracy. The announcement of
opportunity (AO) for the data processing will be issued by ESA late in 2006. The
Gaia Data Processing and Analysis Consortium (DPAC) has been formed recently and
is preparing an answer to this AO. The satellite will downlink around 100 TB of
raw telemetry data over a mission duration of 5--6 years. To achieve its
required accuracy of a few tens of microarcseconds in astrometry, a highly involved
processing of this data is required. In addition to the main astrometric
instrument Gaia will host a radial-velocity spectrometer and two low-resolution
dispersers for multi-colour photometry. All instrument modules share a common
focal plane made of a CCD mosaic about 1 square meter in size and featuring
close to 1 Giga pixels. Each of the various instruments requires a relatively
complex processing while at the same time being interdependent. We describe the
composition and structure of the DPAC and the envisaged overall architecture of
the system. We shall delve further into the core processing - one of the nine,
so-called coordination units comprising the Gaia processing system.
\end{abstract}

\section{Introduction}
This paper is sub-divided in four sections: We give a brief overview of the Gaia
satellite and introduce the Data Processing and Analysis Consortium (DPAC).
Following on from this we describe the overall system architecture for Gaia
processing and finally take a more detailed look at the core processing. 

\section{The Gaia Satellite and Science} 
The Gaia payload consists of three distinct instruments for astrometric,
photometric and spectroscopic measurements, mounted on a single optical bench.
Unlike HST and SIM, which are pointing missions observing a preselected list of
objects, Gaia is a scanning satellite that will repeatedly survey in a
systematic way the whole sky. 
The main performances of Gaia expressed with just a few
numbers are just staggering and account for the vast scientific harvest awaited
from the mission: a complete survey to 20th magnitude of all point sources
amounting to more than one thousand million objects, with an astrometric
accuracy of 12--\pu{25}{$\mu$as} at 15th magnitude and \pu{7}{$\mu$as} for the
few million stars brighter than 13th magnitude; radial velocities down to 17th
magnitude, with an accuracy ranging from 1 to \pu{15}{km\,s$^{-1}$}; multi-epoch
spectrophotometry for every observed source sampling from the visible to the near IR.

Beyond its sheer measurement accuracy, the major strength of Gaia follows from
(i) its capability to perform an all-sky and sensitivity limited absolute
astrometry survey at sub-arcsecond angular resolution, (ii) the unique combination into a single spacecraft of the
three major instruments carrying out nearly contemporaneous observations, (iii)
the huge number of objects and observations allowing to amplify the accuracy on
single objects to large samples with deep statistical significance, a feature
immensely valuable for astrophysics and unique to Gaia.

\section{The Data Processing and Analysis Consortium (DPAC)}
\begin{figure}[Htb]
\begin{center}\mbox{}\\[0mm]
\plotone{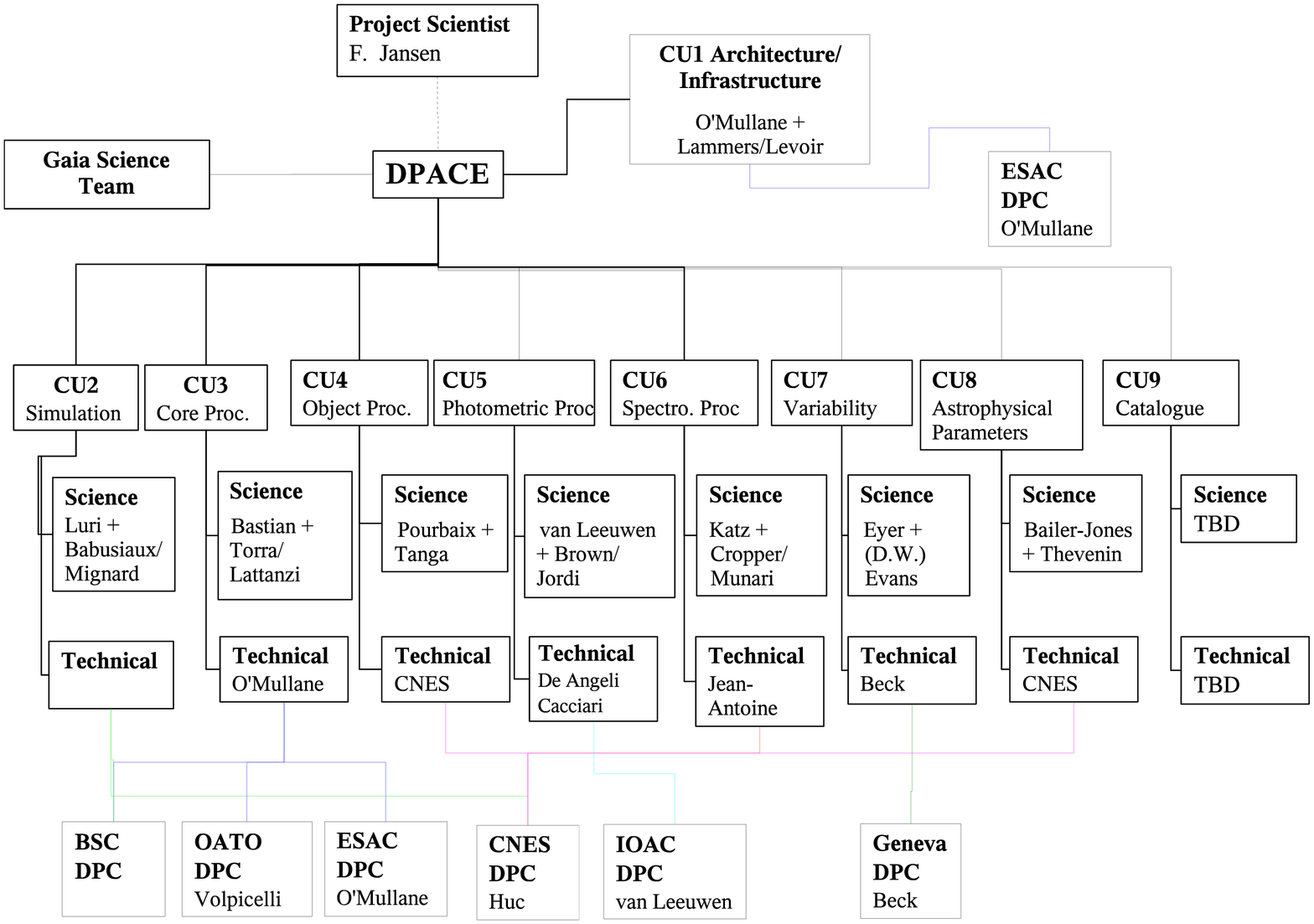}
\epsscale{0.8}
\end{center}
\caption{DPAC Organigram showing DPACE, Coordination Units and Data processing
Centres
\label{fig:dpacorgan}}
\end{figure}

The DPAC has been formed to answer the Announcement of Opportunity (AO) for the
Gaia Processing. The DPAC is formed around a series of ``Coordination Units'' (CU),
themselves sub-divided into ``development units'' (DU). The CUs are supported by
a set of Data Processing Centres (DPC). The overall coordination is performed by
the consortium executive (DPACE). The structure is shown in
\figref{fig:dpacorgan} and described in more detail below. Consider that there are
over 270 scientists and developers currently registered in DPAC who will 
contribute to the scientific processing on Gaia.
 
\subsection{Coordination Units (CU)}
The CUs are small in number, with clearly-defined responsibilities and
interfaces, and their boundaries fit naturally with the main relationships
between tasks and the associated data flow.

There will be several areas of involvement across these boundaries, but in first
instance it is up to the coordination units to ensure that a group of tasks is
prepared and optimised, as well as fully tested and documented, as required by
the project.

The coordination units will have a reasonable amount of autonomy in their
internal organisation and in developing what they consider as the best solution
for their task. However they are constrained by the fact that any such solution
has to meet the requirements and time schedules determined by the Consortium
Executive for the overall data processing. In this respect the data exchange
protocol and the adherence to the data processing development cycles are mandatory to ensure
that every group can access the data it needs in the right format and at the
right moment. While the coordination units are intended to reflect the top level
structure of the data processing, with completely well-defined responsibilities
and commitments to the DPAC, they could for practical reasons be sub-divided
into more manageable components, called development units (DU). This is a more
operational level with a lighter management which will take the responsibility
for the development of a specific part of the software with well defined
boundaries. Not every CU will organise its DUs (if any) in the same way, and how
they interact with the CU level is left to the CU management.

Responsibilities of the coordination units include: (a) establishing development
priorities; (b) procuring, optimizing and testing algorithms; (c) defining and
supervising the development units linked to them. Each coordination unit is
headed by a scientific manager and one or two deputy managers. The CUs will also comprise
software engineers.

\subsection{Data Processing Centres (DPC)}
The development activities of each CU are closely associated with at least one
DPC (Data Processing Centre) where the computer hardware is available to carry
out the actual processing of the data. A technical manager from this centre
belongs to the upper management structure of every CU. The software development
and the preparation and testing of its implementation in a DPC are parallel
activities within every CU and their mutual adequacy must be closely monitored
by the CU manager with his DPC representative. Advancement reports are regularly
presented to the DPAC executive.

\subsection{DPAC Executive (DPACE)\label{ sect:DPACE}}
Our overall organisation gives the CUs much autonomy in the way they handle
their part of the data processing, and the internal organisation and management
structures do not need to be uniform across CUs. However there is a single goal
shared by all the CUs, and they must follow a common schedule and adhere strictly
to many interfaces so that the results produced by one group are available in a
timely manner and may be used efficiently by other groups. 
A variety of standards and conventions, the
content and structure of the MDB (Main Database) and the processing cycles must be
agreed collectively. Therefore in addition to a local management of each CU, the
overall DPAC is coordinated and managed by an Executive Committee, called DPACE
for ``Data Processing and Analysis Consortium Executive''. This overall
management structure of the Consortium deals with all the matters which are not
specific to the internal management of a CU and is meant to make an efficient
interaction between the CUs possible. The DPACE responsibilities are primarily
coordination tasks although it will make important decisions to be implemented
by all CUs which are akin to real management.

\section{Gaia Data Processing Architecture}
\subsection{Approach \label{sect:decomp:approach}}
Any large system is normally broken down into logical components to allow distributed development.
 Gaia data processing is on a very large and highly distributed scale. The approach taken to the 
decomposition has been to identify major parts of the system which may operate relatively 
independently, although practically all parts of the Gaia processing are in fact interdependent
 from the point of view of the data. From a development point of view however, 
a well defined ICD (Interface Control Document) would allow completely decoupled components 
to be developed and even operated in disparate locations. 
The approach is driven by the fact that this is a large system which
will be developed in many countries and by teams of varying 
competence.

Hence at this level of decomposition libraries or infrastructure are not considered to be
 components. At some lower level these components may indeed share libraries and infrastructure 
but this is not a cornerstone for the architecture. 
Only the top level components and their interaction are considered in this decomposition.

\subsection{Logical Components \label{sect:decomp:logic}}
Figure \ref{fig:sysarc} shows the logical components of the 
system and the data flow between them. 

\begin{itemize}
\item{ Mission Control System (MCS)\footnote{The MCS and DDS are
  responsibilities of the Mission Operations Centre (MOC), 
  not part of the DPAC and are included here for completeness.} }
\item{ Data Distribution System (DDS) }
\item{ Initial Data Treatment and First Look (IDT/FL)}
\item{ Simulation (SIM) }
\item{ Intermediate Data Update (IDU) }
\item{ Astrometric Global Iterative Solution (AGIS)}
\item{ Astrometric Verification Unit (AVU) }
\item{ Object Processing (OBJ) }
\item{ Photometric Processing (PHOT) }
\item{ Spectroscopic Processing (SPEC) }
\item{ Variability Processing (VARI) }
\item{ Astrometric Parameters (ASTP) }
\item{ Main Database (MDB) }
\item{ Archive }
\end{itemize}

\subsection{Data Flow \label{sect:decomp:dataflow}}
\begin{figure}[htbp]
\begin{center}
 \plotone{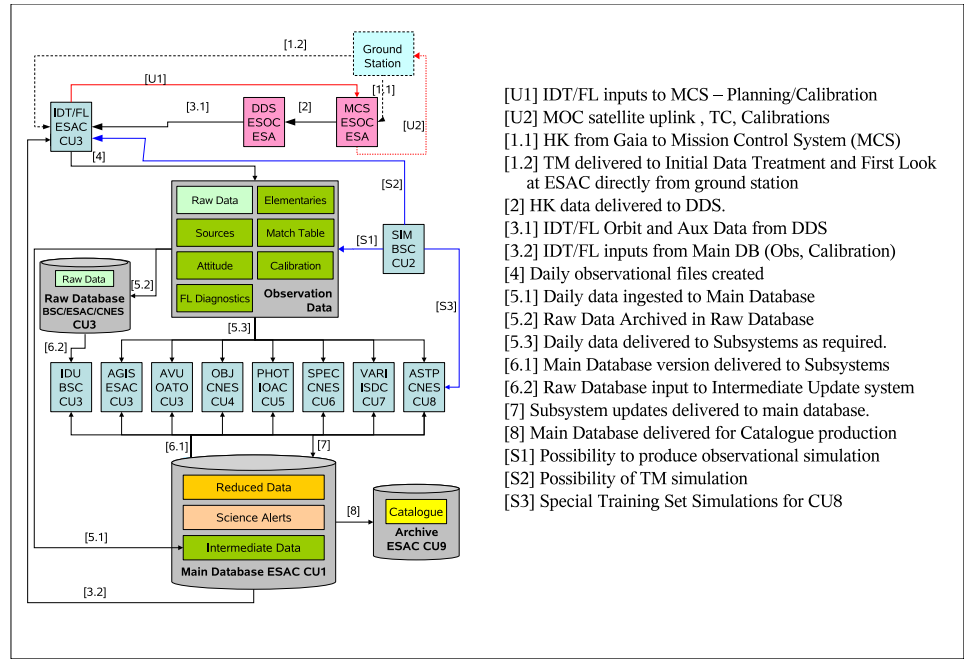}
  \epsscale{0.6}
\end{center}
\caption{Top Level Components and Data Flow for the DPAC}
\label{fig:sysarc}
\end{figure}

Gaia processing is all about data. The data flow is the most important
description of the system and has been under discussion within the community for
some time. The result of these discussions is the data flow scheme depicted in
\figref{fig:sysarc}. The flow lines in \figref{fig:sysarc} are labeled and
these labeled are referred to in the text below. The data flow is divided into
two categories, Near-Realtime and Scientific Processing.

\subsubsection{Near-Realtime dataflow.\label{sect:decomp:nearreal}}
The Near-realtime data flow represents the data flow on a time scale of
approximately 1 or 2 days, corresponding to the activities of the Mission
Operations Ground Segment.

The Mission Operations Centre (MOC) at ESOC receives all telemetry from the Space
Segment [1.1] via the ground stations. The Science Operations Centre (SOC) at
ESAC will receive all telemetry directly from the ground station also [1.2]. The
raw data flow from the satellite is not shown explicitly in the diagram. Over
the nominal mission duration of five years the payload will yield a total
uncompressed data volume of roughly 100 TB. The satellite will have contact with
the ground station once a day for a mean duration of 11 hours. During this
period, or ``pass'', an uncompressed data volume of roughly 50 GB is downlinked
from the satellite via its medium-gain antenna, at a mean rate of about 5~Mbit/s.

\subsubsection{Mission Control System.}

The raw telemetry data received by the ground station will be transmitted to 
the Mission Control System (MCS) at the MOC and to the SOC. Housekeeping data 
will be transmitted to the MCS within one hour after reception
at the ground station.
The MCS will provide an immediate assessment on the spacecraft and instrument 
status through analysis of the housekeeping data.   

\subsubsection{Data Distribution System.} 
All telemetry received by the MOC systems will be ingested into the Data
Distribution System (DDS) [2]. The DDS will also contain data that were
generated on-ground (e.g. orbit data, time correlation data), operations
reports (telecommand history and timeline status), Satellite Databases used by
the MCS and a copy of all telecommands sent to the spacecraft.

\subsubsection{Initial Data Treatment and First Look.}
Science Telemetry is received by the SOC [1.2] for processing by the IDT. Data is also
retrieved from the DDS by the MOC Interface Task at the SOC and passed to the IDT
[3.1]. The IDT processing will decode and decompress the telemetry. It will
also extract higher-level image parameters and provide an initial cross matching
of observations to known sources (or else to new ones created in this step). 
Finally it will provide an initial satellite
attitude at sub-arcsec precision. 

The primary objective of the First Look (FL) is to ensure the scientific health of
Gaia. This information is returned to the MCS [U1]. First Look processing will
carry out a restricted astrometric solution on a dataset from a small number of
great-circle scans.

To perform some of its tasks IDT/FL requires reference data, such as up-to-date
calibration data as well as positions, magnitudes etc.~of 
bright objects that are expected to be
observed by Gaia during the time period to be processed. This data will be made
available to IDT/FL [3.2] from the MDB. FL will also calibrate the current data
set itself and this calibration will be used by IDT.

\subsubsection{Uplink.} 
The telemetry is received by the MCS which does basic system monitoring.
The First Look Diagnostics produced by FL [U1] will indicate if there are
anomalies in the scientific output of the satellite which can be corrected
on-board. After interpretation of the diagnostics, the Flight Control Team 
is informed of the
anomaly, which can be resolved either through immediate commanding or during the
next mission planning cycle.

On a regular basis the MCS will send the prepared command schedule to Gaia [U2], 
taking into account normal planning and inputs from IDT/FL. During a ground 
station pass, immediate commanding is also possible.

\subsubsection{Daily transfers and Raw Database. \label{sect:rawdb}} 
The outputs of IDT/FL are made available to all tasks on a daily basis [4,5] and
ingested into the Main Database [5.1]. The Raw Database will be a repository
for all raw data [5.2]. Copies of the Raw Database are expected at ESAC, BSC and
CNES. Other tasks may retrieve the data according to their requirements [5.3].
Raw data will only be transmitted on a daily basis i.e., it does not form part
of the Main Database and is not foreseen to be sent again later.

Data Processing Centres may produce `science alerts' from the Gaia observations.
Science alerts are sent to the SOC for immediate distribution to the scientific
community and for archiving in the Main Database [7].

\subsection{Scientific Processing\label{sect:decomp:downlink}}

Scientific Processing represents the production of the Gaia data products by the
Data Processing Ground Segment from Intermediate Data. The timescale for each
iteration of this process is much longer than that of the 
near-realtime processing, of the
order of six months or more. It will continue after routine satellite
operations have finished and will culminate in the production of the Gaia
Catalogue. The outputs of processing from each CU will be sent for incorporation
in the Main Database [7]. 

The Main Database is the hub of all data in the Gaia Data processing system. Our
plan is to version this database at regular intervals, probably every six
months. The science processing is in general iterative. Hence each version of the
Main Database is derived from the data in the previous version. By fixing the
versions of the entire dataset at some point in time we avoid tracking
individual object versions for the billions of objects in the database.

\section{Gaia core astrometric processing}
As described above the core processing involves IDT, FL, IDU and AGIS. In this
section we will look at the Astrometric Global Iterative Solution (AGIS) in a
little more detail. The astrometric core solution is the cornerstone of the
data processing since it provides calibration data and attitude solution needed
for all the other treatments, in addition to the astrometric solution of
$\sim100$ million \emph{primary sources}. The main equations to be solved can be
summarized by relating the observed position on a detector to a general
astrometric and instrument model as,
\begin{equation}
   O = S + A + G + C +\epsilon \label{eq:astromsol}
\end{equation}
where
\begin{itemize}
  \item $O$ is the observed one-dimensional location of the source at the
    instant determined by the centroiding algorithm applied to the observed
    photo-electron counts.
  \item $S$ is the astrometric model which for the primary stars should only
    comprise the five astrometric parameters ($\alpha_0, \delta_0, \pi,
    \mu_\alpha, \mu_\delta)$.
  \item $A$ represents the parameters used to model the attitude over a given
    interval of time. They are, for example, the cubic spline coefficients of
    the quaternion describing the orientation of the instrument with respect to the
    celestial reference frame as function of time.
  \item $G$ represents the global parameters such as the PPN parameters or
    other relevant parameters needed to fix the reference frame of the
    observations.
  \item $C$ comprises all the parameters needed for the instrument modelling:
    geometric calibration parameters (both intra- and inter-CCDs), basic angle,
    chromaticity effect and other instrumental offsets. 
  \item $\epsilon$ is the Gaussian white noise which can be estimated from the
    photon counts and centroiding for every observation and used to weigh the
    equations. A test is performed at the end to validate the assumption on
    the noise.
\end{itemize}

Assuming some $10^8$ primary stars, the total number of unknowns for the
astrometric core solution includes some $5\times 10^8$ astrometric parameters,
$\sim 10^8$ attitude parameters, and a few million calibration parameters. The
condition equations connecting the unknowns to the observed data are
intrinsically non-linear, although they generally linearise well at the
sub-arcsec level. Direct solution of the corresponding least-squares problem is
unfeasible, by many orders of magnitude, simply in view of the large number of
unknowns and their strong inter-connectivity, which prevents any useful
decomposition of the problem into manageable parts. The proposed method is based
on the {\em Global Iterative Solution} scheme (ESA 1997, Vol.~3, Ch.~23), which
in the current context is referred to as the {\em Astrometric GIS} (AGIS) since
related methods are adopted for the photometric and spectroscopic processing.
It is necessary to have reasonable starting values for all the unknowns, so as
to be close to the linear regime of the condition equations. These are generally
provided by the Initial Data Treatment.

The idea of AGIS is then quite simple, and consists of the following steps:
\begin{enumerate}
\item
Assuming that the attitude and calibration parameters are known, the
astrometric parameters can be estimated for all the stars. This can be done for
one star at a time, thus comprising a least-squares problem with only 5 unknowns
and of order 1000 observations. Moreover, this part of the solution is extremely
well suited for distributed processing.
\item
Next, assuming that the astrometric parameters and the calibration parameters are
known, it is possible to use the same observations to estimate the attitude.
This can be done for each uninterrupted observation interval at a time. For a
typical interval of one week, the number of unknowns is about 500\,000 and the
number of observations $\sim 2\times 10^7$.  The number of unknowns may seem
rather large for a least-squares problem, but the band-diagonal structure of the
normal equations resulting from the spline fitting makes the memory consumption
and computing time a linear function of the number of unknowns, rather than the
cubic scaling for general least-squares solutions. The problem is thus easily
manageable.
\item
Assuming then that the astrometric and attitude parameters are known, the
calibration parameters can be estimated from the residuals in transit
time and across-scan field angles.
\item
It is necessary to iterate the sequence of steps 1, 2, 3 as many times
as it takes to reach convergence. Once the linear regime has been reached,
the convergence should be geometric, i.e., the errors (and updates) should
decrease roughly by a constant factor in each cycle. Based on simple
considerations of redundancy and the geometry of observations,
a convergence factor of $0.2$--$0.4$ is expected. If a geometric behaviour
is indeed observed, it may be possible to accelerate the convergence by
over-relaxation. The iteration must be driven to a point where the
updates are much smaller than the accuracy aimed at in the resulting
data.
\item
After convergence, the astrometric and attitude parameters refer to an
internally consistent celestial reference frame, but this does not necessarily,
and in general will not, coincide with the International Celestial Reference
System (ICRS). A subset of the primary stars and quasars, with known
positions/proper motions in the ICRS, is therefore analyzed to derive the nine
parameters describing a uniform rotation between the two systems, plus the
apparent streaming motion of quasars due to the cosmological acceleration of the
solar-system barycentre.  The astrometric and attitude parameters are then
transformed into the ICRS by application of a uniform rotation.
\end{enumerate}
It is envisaged that the whole sequence 1--5 is repeated several times during
the processing, initially perhaps every 6 months during the accumulation of more
observations. These repeats are called outer AGIS iterations. Optionally, the
iteration loop 1--3 may also include an estimation of global parameters.

\subsection{AGIS Implementation}
To facilitate the execution of the AGIS algorithms  a data driven approach has been followed.
The notion is that any data should only be read once and passed
to algorithms rather than each algorithm accessing data directly. A process, termed the
'Data Train', sits between the data access layer and the algorithm requiring data. It 
accesses the data and invokes the algorithm thus providing an absolute buffer between
scientific code and data access.
The 'Data Train' is a manifestation of  the 'intermediary pattern' [Gamma 1994] and
has the advantage that different implementations of varying complexity may be 
provided. Hence a scientist may run a simplified AGIS on a laptop for testing which
does not require the full hardware of the data processing centre.

ESAC currently hosts AGIS on a sixteen node cluster of dual processor Dell Xeon
machines. A  6 TB Storage Area Network (SAN) is used to host the Oracle database containing the data. 
The system is entirely written in Java and runs on the 64-bit Sun JDK1.5.  

The current implementation 
is executing with simulation data and reaches convergence within 27 outer iterations for
very noisy input data. The simulated dataset is of 1.1 million sources with five years
of observation amounting to about $10^8$ observations.  Convergence is declared when
the median added to the width of the parallax update histogram is below 1 $\mu$as.
See also [Hernandez 2007].

\section{Conclusion}
Gaia is an ambitious space mission where the instrument and data processing are 
intimately related. An overall distributed data processing architecture has been
outlined. A distributed management
structure is in place to ensure the processing software is built. Rapid development
of key software modules is underway, for example the core astrometric solution has been 
presented in this paper. DPAC has made an excellent start but a difficult road lies  
ahead to achieve the demanding accuracies required by the Gaia mission.


\begin{references}
\reference Hernandez, J. et al. 2007, \adassxvi\ \paperref{P1.05} 
\reference {ESA and FAST and NDAC and TDAC and INCA and Matra Marconi Space and Alenia Spazio 1997 
         {T}he {H}ipparcos and {T}ycho {C}atalogues, SP-1200}
\reference  {Gamma, E. and Helm, R. and Johnson, R. and Vlissides, J. 1994, Addison-Wesley, {D}esign {P}atterns: {E}lements of {R}eusable {O}bject-{O}riented {S}oftware }
\end{references}
\end{document}